\documentclass[onecolumn,12pt]{IEEEtran}
\usepackage{amsmath}
\usepackage{amsthm}
\usepackage{amsfonts}
\usepackage{amssymb}
\usepackage{mathrsfs}
\usepackage{cases}
\usepackage{cite}
\usepackage{graphicx}
\usepackage{mathrsfs}
\usepackage{subfigure}
\usepackage{epstopdf}
\usepackage{color}

\usepackage{enumitem}
\usepackage{setspace}
\usepackage{bm}
\usepackage{caption}
\usepackage{algorithm}
\usepackage{algorithmic}
\usepackage{multirow}
\usepackage{tabularx}
\usepackage{makecell}

\theoremstyle{remark}
\newtheorem{theorem}{\hspace{1em}Theorem}

\begin{document}
\baselineskip 4ex

\title{\LARGE Fast Detection of Burst Jamming for Delay-Sensitive Internet-of-Things Applications
% \thanks{
% The work}
\thanks{
S.-D. Wang and H.-M. Wang are with the School of Information and Communication Engineering, and also with the Ministry of Education Key Lab for Intelligent Networks and Network Security, Xi’an Jiaotong University, Xi’an, 710049, Shaanxi, China (e-mail: xjtuwsd@stu.xjtu.edu.cn; xjbswhm@gmail.com).
}
\thanks{
P. Liu is with the Wireless Technology Lab, 2012 Labs, Huawei Technologies, Shenzhen, China (e-mail: jeremy.liupeng@huawei.com).
}
\author{Shao-Di Wang, Hui-Ming Wang,~\IEEEmembership{Senior Member,~IEEE}, and Peng Liu}}
\maketitle

%\pagenumbering{gobble}

\begin{abstract} 

	In this paper, we investigate the design of a burst jamming detection method for delay-sensitive Internet-of-Things (IoT) applications. In order to obtain a timely detection of burst jamming, we propose an online principal direction anomaly detection (OPDAD) method. We consider the one-ring scatter channel model, where the base station equipped with a large number of antennas is elevated at a high altitude. In this case, since the angular spread of the legitimate IoT transmitter or the jammer is restricted within a narrow region, there is a distinct difference of the principal direction of the signal space between the jamming attack and the normal state. Most of existing binary hypothesis test based works cannot apply to detect burst jamming, because the attackers’ target time window does not match with the legitimate transmission. Unlike existing statistical features based batching methods, the proposed OPDAD method adopts an online iterative processing mode, which can quickly detect the exact attack time block instance by analyzing the newly coming signal. In addition, our detection method does not rely on the prior knowledge of the attacker, because it only cares the abrupt change in the principal direction of the signal space. Moreover, based on the high spatial resolution and the narrow angular spread, we provide the convergence rate estimate and derive a nearly optimal finite sample error bound for the proposed OPDAD method. Numerical results show the excellent real time capability and detection performance of our proposed method.

\end{abstract}
\begin{IEEEkeywords}
Physical layer security, burst jamming, principal direction, online anomaly detection, delay-sensitive.
\end{IEEEkeywords}

\section{Introduction}
\label{Sec:Introduction}

Delay-sensitive Internet-of-Things (IoT) applications have been drawing increasing attention lately. The most characterizing feature of such applications is that messages are required to be transferred in real-time. Delay-sensitive IoT applications are often associated with critical human tasks [1]-[4]. For instance, in various kinds of IoT-based disaster detecting system, the detecting information must be transferred to the decision making center as soon as possible for disaster prevention. Thus, having a high security level is a crucial requirement for delay-sensitive IoT applications [5], [6].

However, delay-sensitive IoT applications are vulnerable to jamming attacks due to the nature of their wireless operating media [7]. In delay-sensitive IoT applications, there is always a message delivery deadline beyond which the message is considered to be useless [8]. Jamming attacks can incur a large transmission delay, which lead to missing the deadline and significantly impact the regular operations of delay-sensitive IoT applications [9], [10]. Compared with constant jamming, i.e., a jammer constantly emits the jamming signals, burst jamming is likely to become even more threatening to delay-sensitive IoT applications because of its energy-efficient and stealthy model [11]. 
In burst jamming, on one hand, the attacker alternates between sleeping and jamming modes to save energy and has higher jamming power due to the short active period, which leads to a rapid increase in the number of dropped/retransmitted packets, resulting in a large transmission delay. Because the attackers aim to increase the transmission delay rather than reducing the throughput, burst jamming is generally more efficient. On the other hand, burst jamming lowers the risk of detection because it seems quite random, and thus a detector might not be able to distinguish whether signals on the channel are from a IoT transmitter or a jammer. Note that constant jamming can easily be detected and identified in delay-sensitive IoT applications with high demand of security, burst jamming is more covert and difficult to perceive.

In order to ensure countermeasures can be timely taken, e.g., adaptive array beamforming [12], interference cancellation techniques [13], interference alignment techniques [14], a timely detection of burst jamming is a critical issue to be addressed for delay-sensitive IoT applications.

\subsection{Related Literatures}

Jamming detection has received considerable attentions, and various detection methods have been studied and proposed. Generally, jamming detection can be performed by a \emph{statistic feature (SF) recoginition and classification approach}, different detection methods use different statistics for decision making. These statistic features can be classified into the following two categories: \emph{1) statistical features of physical layer [15]-[19]}; \emph{2) statistical features of upper layer [20]-[24]}.

\emph{1) Statistical features of physical layer [15]-[19]:} In [15], the authors proposed an energy detector (ED) based method. This method relies on the fact that the received energy is quite different than a predesigned threshold in the occurrence of jamming attacks. In [16], the subspace dimension (SD) based method was proposed to detect a structured signal from unknown jamming attacks by extracting the subspace dimension of signal covariance matrix. In [17], the authors proposed to exploit the variance and channel state information (CSI) based methods to detect jamming or illegitimate wireless network access interferes. The detection framework focuses on distinguishing between legitimate and illegitimate transmissions and the nature of illegitimate transmissions with a quaternary hypotheses test. In [18], the authors derived several detectors for adaptive detection in a generalized multivariate analysis of variance signal model with structured interference. The maximal invariant statistic (MIS) was utilized to design the suitable detectors which can possess the constant false alarm rate property. In [19], the authors studied channel-aware decision fusion in a wireless sensor network with interfering sensors, and developed five sub-optimal fusion rules by exploiting a second-order characterization (SOC) of the received vector to detect the attacks.

\emph{2) Statistical features of upper layer [20]-[24]:} In [20], the effective channel utilization (ECU) metric is computed and used as a statistic to detect jamming attacks. ECU is a widely used metric that measures the channel utilization in a wireless network. The occurrence of jamming attack was claimed if the ECU was larger than a predesigned threshold. In addition, the packet delivery ratio (PDR) based detection method is widely used in [21]-[23], a monitoring node keeps track of the percentage of transmission collisions in a wireless network, and a jamming attack is detected when the percentage exceeds a certain threshold. Other works such as [24], the authors proposed to exploit the network throughput (NT) based feature to detect the jamming attack.

Although there have been many research works proposing jamming detection, existing methods are not suitable for quick detection of burst jamming in delay-sensitive IoT applications. The primary causes can be summarized as follows:

\emph{1) Burst jamming detection:} In existing works, the problem of jamming detection was modeled as a binary hypothesis test problem, i.e., jamming either is everlasting all the time or does not exist at all. It is worth noting that an implicit assumption is that the attackers’ target time window perfectly matches with the legitimate transmission. However, in burst jamming, the start time of the attack is usually unknown, and it is likely that it starts at the middle of the legitimate transmission. Thus, such a binary hypothesis test model as in most of existing methods cannot apply to detect burst jamming.

\emph{2) Real time detection:} Delay-sensitive IoT applications tend to focus more on delay performance than throughput. Until now, the problem of real time detection has not yet been fully taken into consideration. Because the existing detection methods are mostly implemented in a batch manner, which need to store and manipulate a large number of observation signals for analysis and processing. However, such a batch mode not only requires massive storage and computing resources, but also affects the real time performance of jamming detection. So, how to quickly detect jamming attacks against delay-sensitive message delivery is very urgent.

\emph{3) Imperfect prior information:} Most of the related works rely on the prior knowledge of the attacker to choose a decision threshold to distinguish the jamming attack from the normal state, which is unrealistic. Because the adversary should not cooperate with the legitimate system, it is not easy to obtain such statistics, especially that of burst jamming, such as the attack power and the start time of the jammer.

\subsection{Motivations and Contributions}

Aiming at detecting the occurrence of burst jamming as quickly as possible once it starts in delay-sensitive IoT applications, in this paper, we propose an online principal direction anomaly detection (OPDAD) method. Our OPDAD method is motivated by the fact that the principal direction of the signal space (hereinafter referred to as principal direction) will be changed with high probability if burst jamming happens. This because the facing scatters are different, the legitimate IoT transmitter and the attacker with different locations will result in different angle of arrivals (AoAs) at the receiver, and each element in the channel of the legitimate IoT transmitters or the attackers indicates the distribution of gain in a specific direction. As a result, there is a distinct difference of the principal direction between the jamming attack and the normal state. Besides, note that the interference signal space is determined by interference channel space [25]. Based on these characteristics, the principal direction as an available physical feature can be exploited to distinguish the normal case and burst jamming with low probability of false alarm. 

In the proposed OPDAD method, we first extract the principal direction incrementally by processing the received signals one by one in a real time manner. Then, centroid-based clustering is used to cluster the received signals into two classes, thereby determining whether burst jamming exists. To the best of our knowledge, we are the first to study burst jamming detection in a real time manner. The superiorities of our proposed detection method compared with the typical methods are listed in Table I. Note that in the considered delay-sensitive case in this paper, we mainly focus on the statistical features of physical layer based methods. This is because the statistical features of upper layer based methods will cause an intolerable delay. Specifically, the proposed detection method can tackle the above three major issues as follows:

\newcommand{\tabincell}[2]{\begin{tabular}{@{}#1@{}}#2\end{tabular}} 
\begin{table*}[t]\small
\centering
\caption{\small COMPARISON BETWEEN DIFFERENT JAMMING DETECTION METHODS}
\begin{tabular}{|c|c|c|c|c|c|c|c|} 
\hline
\hline
Category & Literature & Statistics & {\makecell[c]{Burst jamming \\ detection}} & {\makecell[c]{Delay \\sensitive}}& {\makecell[c]{Without \\attack strategy}} & {\makecell[c]{Without \\attacker data}} & {\makecell[c]{Large-scale \\scenario}}
\\
\hline
\multirow{6}*{\tabincell{c}{Statistical\\ features of \\physical layer}} &\textbf {\makecell[c]{proposed \\ method}}  & \textbf{\makecell[c]{principal \\direction}} & \checkmark & \checkmark & \checkmark & \checkmark & \checkmark\\
\cline{2-8} & [15] & ED-based &  &  & \checkmark & \checkmark & \checkmark\\
\cline{2-8} & [16] & SD-based  &  &  & \checkmark & \checkmark &\\
\cline{2-8} & [17] & CSI-based &  &  &  &\checkmark  &\checkmark \\
\cline{2-8} & [18] & MIS-based &  &  &  &\checkmark  & \checkmark\\
\cline{2-8} & [19] & SOC-based &  &  &  &\checkmark  & \\

\hline
\multirow{3}*{\tabincell{c}{Statistical\\ features of \\upper layer}} & [20] & ECU-based &  &  &\checkmark  &  & \\
\cline{2-8} & [21]-[23] & {\makecell[c]{PDR-based}} &\checkmark  &  &\checkmark  &  & \\
\cline{2-8} & [24] & {\makecell[c]{NT-based}}  &  &  &\checkmark  &  &\checkmark \\

\hline
\hline
\end{tabular}
\end{table*}

\emph{1)} The proposed OPDAD method can detect whether burst jamming exists based on the principal direction anomaly. Once the newly coming signal is jammed, the principal direction will be changed with high probability. If it indicates that no change has occurred, then the detector moves to the next time point until the detection result indicates that a change has occurred.

\emph{2)} Our OPDAD method is capable of detecting jamming attacks in a real time manner. It adopts an online iterative processing mode that takes one observation at a time without having to re-explore all previously available observations. Thus it can quickly detect the exact attack time block instance by analyzing the newly coming signal. Compared with the existing batch methods, the proposed method can reduce the complexity and latency.

\emph{3)} Unlike the existing methods, the initial iteration of the proposed detection method does not require any prior knowledge about either the statistical or the time-variant features of the received signals, this is because the centroid-based clustering does not need any prior information related to the jammers, it eliminates the difficulty of detection threshold determination.

\emph{Organization:} In Section II, we introduce the system model with burst jamming and analyze the limitations of existing works under binary hypothesis test framework. In Section III, we first introduce the detection principle of the proposed OPDAD method, and then present the complete detection framework of our OPDAD method, including the principal direction tracking and the centroid-based clustering. In Section IV, we provide the convergence results and complexity analysis of the proposed detection method. Numerical results are presented in Section V. Finally, Section VI concludes the paper. Table II lists the main acronyms used in this paper.

\begin{table*}[t]\small

\centering
\caption{\small ACRONYMS}
\begin{tabular}{|c|c|} 
\hline
\hline
Acronym & Meaning \\
\hline
IoT & Internet-of-Things  \\
\hline
ED & energy detector  \\
\hline
SD & subspace dimension  \\
\hline
CSI & channel state information   \\
\hline
MIS & maximal invariant statistic  \\
\hline
SOC & second-order characterization  \\
\hline
PDR & packet delivery ratio \\
\hline
NT & network throughput  \\
\hline
OPDAD & online principal direction anomaly detection  \\
\hline
AoA & angle of arrival  \\
\hline
BS & base station  \\
\hline
GLRT & generalized likelihood ratio test  \\
\hline
DMF & direct matrix factorization  \\

\hline
\hline
\end{tabular}
\end{table*}

\emph{Notations:} $(\cdot)^{T}$, and $(\cdot)^{H}$ denote transpose and conjugate transpose, respectively. $\mathbb{V}\rm{ar}(\cdot)$ and $\mathbb{E}(\cdot)$ denote the mathematical variance and expectation, respectively. $\mathbb{E}\{{x| {\mathcal A}}\}$ denotes the expectation of random variable $x$ over event $\mathcal A$. $\pmb{I}_M$ denotes a $M\times M$ identity matrix. Diagonal matrix is denoted by $\rm{diag}(\cdot)$. $|\cdot|$ and $||\cdot||$ denote the absolute value and the $l_2$ norm, respectively. $\lfloor x \rfloor $ denotes the floor function, i.e., the largest integer $ \le x$, and $\lceil x \rceil $ denotes the ceil function, i.e., the smallest integer $ \ge x$. Let $x \wedge y$ denotes the minimum value between $x$ and $y$, and $x \vee y$ denotes the maximum value between $x$ and $y$. $x\rightarrow y$ denotes that $x$ is close to $y$. $\Re(\cdot)$ and $\Im(\cdot)$ denote real and imaginary parts of a complex number. $\mathbb{C}^{N\times M}$ and $\mathbb{R}^{N\times M}$ denote the spaces of all $N\times M$ matrices with complex-valued and real-valued elements, respectively. $\mathbb{CN}(\pmb{\mu},\pmb{\bm R} )$ and $\mathbb{N}(\pmb{\mu},\pmb{\bm R} )$ denote the distributions of complex and real Gaussian random vectors, respectively, with mean $\pmb{\mu}$ and covariance matrix $\pmb{\bm R}$. \\

\section{System Model and Problem Statement}

\subsection{System Model}

We consider a wireless IoT disaster monitoring system with delay requirement depicted in Fig. 1, where $K$ single-antenna legitimate IoT transmitters simultaneously transmit signals to an $M$-antenna base station (BS) in the presence of $N$ single-antenna jammers. To monitor the condition of the disaster detecting system and make real-time decisions, the BS must receive the detecting information from the IoT transmitters in time. Thus, a swift alarm on jamming attack is urgently required.

\begin{figure}[t]
 	\centering
 	\includegraphics[width=6.2 in]{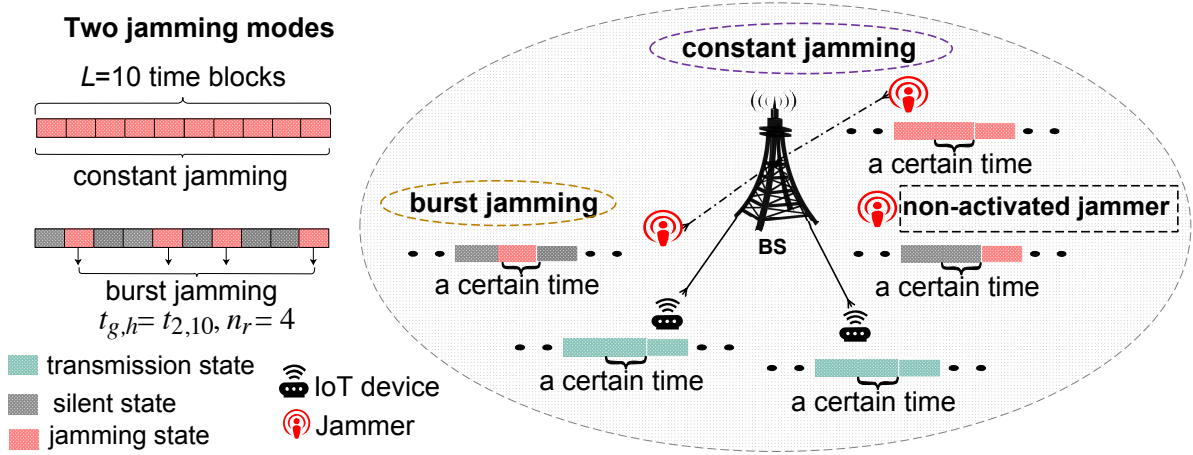}
 	\caption{ System model.}
 	
 \end{figure}

In this paper, the one-ring scatter channel model is considered [26], [27], where each legitimate IoT transmitter or jammer is surrounded by a ring of scatters. In practice, the BS equipped with a large number of antennas is elevated at a high altitude, such that there are few surrounding scatters. In this case, the one-ring model is a reasonable channel model. We use $\bm{h}_{B,k}\sim\mathbb{C}\mathbb{N}\left( \bm{0},\bm{R}_{B,k} \right)$ and $\bm{h}_{J,n} \sim\mathbb{C}\mathbb{N}\left( \bm{0},\bm{R}_{J,n} \right)$ to denote the channel from the $k$th legitimate IoT transmitter and the $n$th attacker to the BS, respectively. According to [27], the covariance matrix of $\bm{R}_{B,k} $ can be calculated by ${\left[ {{\bm R_{B,k}}} \right]_{p,q}} = \frac{1}{{2{\Delta _{B,k}}}}\int_{{{\bar \theta }_{B,k}} - {\Delta _{B,k}}}^{{{\bar \theta }_{B,k}} + {\Delta _{B,k}}} {{e^{ - j\left( {p - q} \right)\pi \sin \left( \theta  \right)}}} d\theta $, where ${\left[ {{\bm R_{B,k}}} \right]_{p,q}}$ denotes the entry in the $p$th row, $q$th column of ${\bm R_{B,k}}$ with $p,q \in \left\{ {1,2, \cdots ,M} \right\}$. ${{\bar \theta }_{B,k}}$ is the mean AoA of clusters surrounding the BS and ${\Delta _{B,k}}$ is the angular spread of the $k$th IoT transmitter’s channel. $\bm{R}_{J,n}$ can be expressed by the similar form with ${{\bar \theta }_{J,n}}$ and ${\Delta _{J,n}}$. Furthermore, We divide the whole transmission process into $L$ time blocks with equal and fixed length, and assume that all the channel coeffificients are independent and identical distributed (i.i.d.) over different time blocks.

In the considered case in this paper, we focus on the design of burst jamming detection, wherein the jammers activate on a specific time block in a sudden and sporadic manner. From the perspective of the BS, the activation patterns of burst jamming seem random due to the unknown strategy of the attackers. We use $n_r$ to denote the number of burst jamming attacks during the target time window $t_{g,h}$, namely the time period from the $g$th to the $h$th time block. In order to illustrate the difference between burst jamming and constant jamming, an example is given in Fig. 1 over $L=10$ time blocks. In burst jamming, the attack target time window is $t_{2,10}$, and the number of burst attacks is $n_r=4$. When it comes to constant jamming, the jammers keep active during all the time. Note that the CSI of the attackers are unknown to the BS. In the following subsection, we point out the limitations of binary hypothesis test framework in existing detection methods.

\subsection{Limitations of Binary Hypothesis Test Framework}

During the attack, the zero-mean unit power signal ${x_{k,l}}$ and the jamming signals ${s_{n,l}}$ are simultaneously transmitted by the $k$th legitimate IoT transmitter and $n$th jammer in the $l$th time block, $l \in \left\{ {1,2, \cdots ,L} \right\}$, respectively. No CSI is available to the jammers, and thus it would be reasonable for the jammers to transmit a noise-like jamming signals with equally distributed power [28]. Therefore, in this paper, the jamming signals are assumed to be i.i.d. Gaussian random variables with zero mean and unit variance, which is independent to the signal ${x_{k,l}}$.

The existing methods under the binary hypothesis test framework collect all sample observations before making a decision, i.e., the received signal at the BS over $L$ time blocks, denoted by ${\bm{y}_{B}}$, can be modeled as
\begin{align}
\label{DetectionModel}
\bm{y}_{B}=\left\{
\begin{aligned}
&{\underbrace {\sum\limits_{k = 1}^{{K}} {\sqrt {{P_{{U_k}}}} \bm{H}_{B,k}^{}} \bm x_{k}^{}}_{\rm {from\ the\ IoT\ transmitters}} + {\bm{w}_{B}}}, && \rm {no\ jamming},\\
&{\underbrace {\sum\limits_{k = 1}^K {\sqrt {{P_{{U_k}}}} \bm{H}_{B,k}^{}} \bm x_{k}^{}}_{\rm {from\ the\ IoT\ transmitters}} + \underbrace {\sum\limits_{n = 1}^N {\sqrt {{P_{{J_n}}}} \bm{H}_{J,n}^{}{\rm diag}({\bm \alpha_{\theta,n}}){ \bm s_{n} }}}_{\rm {from\ the\ jammers}}+ {\bm{w}_{B}}}, && \rm {jamming},
\end{aligned}\right.
\end{align}
where ${P_{{U_k}}}$ is the transmit power of the $k$th IoT transmitter, $\bm{H}_{B,k} \triangleq [\bm{h}_{B,k,1},\cdots, \bm{h}_{B,k,l},\cdots,\bm{h}_{B,k,L} ]\in {{\mathbb C}^{M \times L}}$, $\bm{h}_{B,k,l}$ denotes the channel $\bm{h}_{B,k}$ in the $l$th time block and $\bm x_{k} \triangleq [x_{k,1},x_{k,2},\cdots, x_{k,L} ]^T \in {{\mathbb C}^{L \times 1}}$. ${P_{{J_n}}}$ is the jamming power of the $n$th jammer, $\bm{H}_{J,n} \triangleq [\bm{h}_{J,n,1},\cdots,\bm{h}_{J,n,l},\cdots,\bm{h}_{J,n,L} ]\in {{\mathbb C}^{M \times L}}$, $\bm{h}_{J,n,l}$ denotes the channel $\bm{h}_{J,n}$ in the $l$th time block and $\bm s_{n} \triangleq [s_{n,1},s_{n,2},\cdots, s_{n,L} ]^T \in {{\mathbb C}^{L \times 1}}$, and ${\bm{w}_{B}}\sim\mathbb{C}\mathbb{N}\left( \bm{0},\sigma _B^2{\bm{I}_{M}} \right)$ is the the additive white Gaussian noise (AWGN) at the BS. ${\bm\alpha_{\theta,n}}\triangleq  [{\alpha_{\theta,n,1},\cdots , \alpha_{\theta,n,l}, \cdots ,\alpha_{\theta,n,L}}]^T $, $\theta \in \{C,B\}$, is the attack activity indicator for $n$th jammer, i.e., ${\alpha_{\theta,n,l}}$ is set to one (zero) if the $n$th jammer is active (inactive) in the $l$th time block, where $\theta =C$ indicates constant jamming, and $\theta =B$ stands for burst jamming. For example, as shown in Fig. 1, for $n$th jammer over $L=10$ time blocks, the attack activity pattern $\bm \alpha_{C,n}$ for constant jamming is set to $[1,1,1,1,1,1,1,1,1,1]^T$, and $\bm \alpha_{B,n}$ for burst jamming is set to $[0,1,0,0,1,0,1,0,0,1]^T$. To complete the testing formulation, we can model the change of the $l$th element of $\bm \alpha_{B,n}$ across time as a Markov chain [29] to characterize the time-variation of $\bm \alpha_{B,n}$.

In such a binary hypothesis test problem, the following two hypothesises were taken into consideration, $\mathcal{H}_0$: jamming does not exist; $\mathcal{H}_1$: jamming exists. Due to the CSI of the attackers is unknown to the BS, in order to solve this hypothesis test problem, existing methods follow the framework of generalized likelihood ratio test (GLRT). Specifically, in order to design an effective threshold, the detection feature, denoted by $\tau$, needs to be extracted from the observation signals. Take the ED-based method [15] for example, the received signal by the $m$th antenna at the BS, denoted by ${y}_{B}\left[ m \right]$, for $m = 1,2, \cdots ,M$, and the detection feature based on energy, denoted by $e_m$, can be written as $\tau = e_m \triangleq {\mathbb V}{\rm{ar}}\{{y}_{B}\left[ m \right]\}$. However, the above detection framework could not detect burst jamming timely, there are three main and essential reasons about it: 

1) In existing binary hypothesis test based works, it was implicitly assumed that jamming either is everlasting all the time or does not exist at all, e.g., the attack target time window should be $t_{1,10}$ and the number of attacks $n_r$ should be 10 over $L=10$ time blocks. But in burst jamming, the start time and the number of attacks are usually unknown to the BS caused by $\bm \alpha_{B,n}$. The attackers’ target time window $t_{2,10}$ does not match with the legitimate transmission and the number of attacks $n_r$ is random. 

2) The above detection process needs to collect all the received signals before making a decision, i.e., using the sample correlation matrix of $L$ observations $\bm R_L \triangleq {\bm{y}_{B}} \bm{y}_{B}^H$, and extract the signal features in a batch manner. 
Take the SD-based method [16] for example, subspace dimension $ {\rm rank}(\bm R_L)$ is difficult to detect the exact attack time block instance timely. We use $\bm y_{B,l}$ to denote the received signal at the BS in the $l$th time block. In fact, the detector of burst jamming should work in an online manner, i.e., the detector makes a decision at each time block when a new sample observation $\bm y_{B,l}$ is obtained. Only in this way can we detect the exact attack time block instance.

3) Since the legitimate receiver has no knowledge of the attackers’ target time window $t_{g,h}$ caused by $\bm \alpha_{B,n}$, the binary hypothesis test model as in most of existing methods are no longer applicable. In addition, due to the multiple switching from silence to activism, it is difficult to calculate the maximum likelihood estimates $\hat \tau$ of $\tau$, which further leads to difficulties in determining the GLRT detector, i.e., $\mathcal L(\bm{y}_{B})=f(\bm{y}_{B}, \hat \tau;\mathcal{H}_1)/f(\bm{y}_{B};\mathcal{H}_0)$. As a result, it is difficult to compute the distributions of the sufficient test statistics $T_G$, not to mention the detection threshold $\eta_G$ for a given false alarm ${p_{fa}} \triangleq P\{ T_G>\eta_G| \mathcal{H}_0 \}$.

In order to deal with these problems, we propose a burst jamming detection method through the abrupt change of the principal direction, which can detect the occurrence of burst jamming timely by processing the received signals one by one. 

\section{OPDAD method for burst jamming detection}

In this section, we first introduce the detection principle of the proposed OPDAD method, and then present the complete detection framework of our OPDAD method.

\subsection{Detection Principle of OPDAD}

Our OPDAD method mainly relies on the following facts that, in practice, the BS equipped with a large number of antennas is elevated at a high altitude, such that there are few surrounding scatters. In this case, the angular spread of the $k$th IoT transmitter’s channel is restricted within a narrow region $\left[ {{{\bar \theta }_{B,k}} - {\Delta _{B,k}},{{\bar \theta }_{B,k}} + {\Delta _{B,k}}} \right]$. Similarly, the angular spread of the $n$th attacker’s channel is restricted within a narrow region $\left[ {{{\bar \theta }_{J,n}} - {\Delta _{J,n}},{{\bar \theta }_{J,n}} + {\Delta _{J,n}}} \right]$. In addition, the overlapping area between $\left[ {{{\bar \theta }_{B,k}} - {\Delta _{B,k}},{{\bar \theta }_{B,k}} + {\Delta _{B,k}}} \right]$ and $\left[ {{{\bar \theta }_{J,n}} - {\Delta _{J,n}},{{\bar \theta }_{J,n}} + {\Delta _{J,n}}} \right]$ is quite small. This is because the facing scatters are different, the legitimate IoT transmitter and the attacker with different locations will result in different AoAs at the BS, and each element in $\bm{h}_{B,k}$ or $\bm{h}_{J,n}$ indicates the distribution of gain in a specific direction. As a result, there is a distinct difference of the principal direction between the jamming attack and the normal state. Note that the interference signal space is determined by interference channel space [30]. 

Based on the high spatial resolution and the narrow angular spread, we propose to exploit the principal direction to detect burst jamming. The basic principle of the OPDAD method is shown in Fig. 2. Besides, in the considered case in this paper, the attackers are not active during the channel training phase. The statistics of the legitimate IoT transmitters’ channel can become known by averaging these channel observations across several channel coherent timing blocks. Thus, the principal direction of the signal space from the legitimate system can be obtained as a priori information to reduce false alarm probability.

\begin{figure}[t]
 	\centering
 	\includegraphics[width=5.5 in]{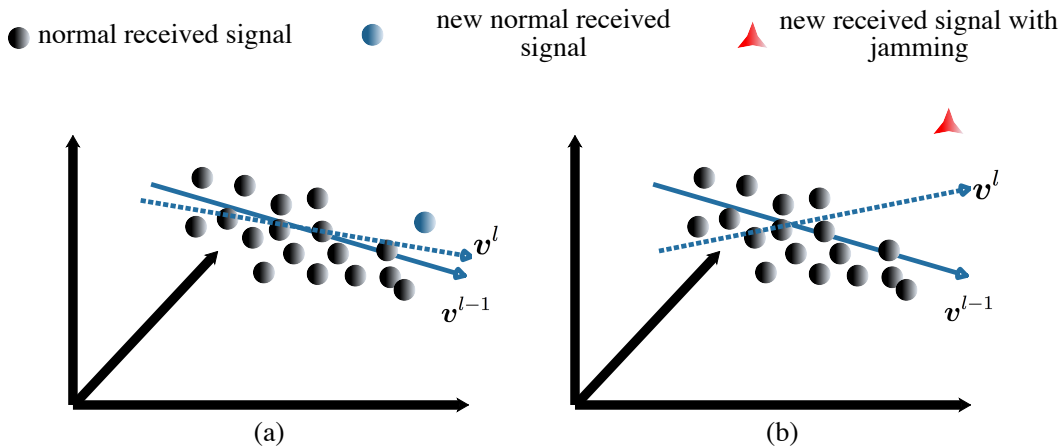}
 	\caption{ Basic principle of the proposed OPDAD method.}
 	\label{f12}
 \end{figure}

\subsection{Real Time Detection Framework of OPDAD}

Our goal is to quickly and accurately detect whether a burst jamming lauches immediately when it happens. We emphasize that the proposed OPDAD method works in an online manner thanks to the principal direction tracking, which can extract the principal direction incrementally by processing the received signals one by one. Specifically, the OPDAD method can be efficiently carried out by the following two stages, including the principal direction tracking and the centroid-based clustering. 

\emph{1) Principal Direction Tracking}

In the proposed OPDAD method, we first need to do is to extract the principal direction as an available detection feature. The received signal at the BS in the $l$th time block, denoted by ${\bm{y}_{B,l}}$, is a random vector in $\mathbb{C}^M$ with mean zero and unknown covariance matrix $\bm Q$. With the aid of the principal direction tracking in the real domain, the extraction of the principal direction can be processed more efficiently. Such transformation in the real domain not only makes it more convenient to provide the convergence speed estimate for the principal direction extraction but also benefits the follow-up centroid-based clustering to determine feasible thresholds to distinguish normal operation and the attacks. Let $\bm y_{B,l} = \Re(\bm y_{B,l})+i\Im(\bm y_{B,l})=\bm m_{B,l}+i\bm n_{B,l}$, and define ${\bm{y}_{R,B,l}} \triangleq {\left( {\bm m_{B,l},\bm n_{B,l}} \right)^T}$, where $i$ is the imaginary unit, we have $\bm y_{B,l}^H {\bm y_{B,l}} = {\left( {\bm {m}_{B,l} - i\bm {n}_{B,l}} \right)^T}\left( {\bm {m}_{B,l}+ i\bm {n}_{B,l}} \right) = \bm {y}_{R,B,l}^T\bm {y}_{R,B,l}$. We use $\bm t_l$ to denote the principal eigenvector of $\bm Q$ and let $\bm Q=\bm A+i\bm B$. Let $\bm t_l= \Re(\bm {t}_l)+i\Im(\bm {t}_l)={\bm {t}}_{R,l}+i\bm{t}_{I,l}$, then we have $\bm{t}_l^H\bm Q \bm t_l = \bm {t}_{R,l}^T\bm A\bm {t}_{R,l} - \bm {t}_{R,l}^T\bm B\bm {t}_{I,l}^{} + \bm {t}_{I,l}^T\bm B\bm {t}_{R,l}^{} + \bm {t}_{I,l}^T\bm A\bm {t}_{I,l}^{}=\bm {v}_l^T\bm \Xi \bm v_l$, where $\bm v_l \triangleq {\left( {\bm t_{R,l},\bm t_{I,l}} \right)^T}$ denotes the principal direction of $\bm \Xi \triangleq$$\begin{matrix}
   \left[ {\matrix
   { \bm A } & {  -\bm B }  \\ 
   {  \bm B} & { \bm A}  \\ 
 \endmatrix } \right]
\end{matrix}$, and $\bm \Xi$ can be regarded as the covariance matrix of real variable $\bm{y}_{R,B,l}$. As a result, we reformulate the complex signal ${\bm{y}_{B,l}}$ in ${{\mathbb C}^{M }}$ with principal direction $\bm t_l$ as an equivalent real variable ${\bm{y}_{R,B,l}}$ in ${{\mathbb R}^{2M }}$ with the principal direction $\bm v_l$.

Based on the new sample observation $\bm{y}_{R,B,l}$, the proposed OPDAD method can extract the principal direction $\bm{v}_{l}$ with the assistance of the principal direction tracking in time. We use $\hat{\bm v_l}$ to denote the estimation result of the principal direction through the principal direction tracking, which can be obtained by maximizing the Rayleigh quotient $G\left( \bm {v}_l  \right) = \frac{{{\bm {v}_l ^T}{\mathbb E}\left\{ {\bm {y}_{R,B,l}^{}\bm {y}_{R,B,l}^T} \right\}\bm {v}_l }}{{\left\| {{\bm {v}_l ^T}\bm {v}_l } \right\|}}$. We measure the quality of the solution $\hat{\bm v_l}$ at time block $l$ using the potential function ${\Psi _l} =1-( \hat{\bm v}_l^T\bm v_l^\odot)/||\hat{\bm v_l}||^2$, where $\bm v_l^\odot$ denotes the true principal direction which can be extracted from the brute force approach, e.g., the direct matrix factorization (DMF) based method [31]. This quantity ranges from 0 to 1, and we are interested in the rate at which it approaches zero. In the principal direction tracking, we require that stepsize $\beta $ be proportional to $1/l$ and that $||\bm {v}_{l}||$ be bounded. Since the $||\bm {v}_{l}||$ correspond to coordinate directions, each update changes just one coordinate of the feature direction. Then, the gradient can be computed by
\begin{align}
\nabla G\left( \bm {v}_l  \right) = \frac{2}{{{{\left\| \bm {v}_l  \right\|}^2}}}\left( {{\mathbb E}\left\{ {\bm {y}_{R,B,l}^{}\bm {y}_{R,B,l}^T} \right\} - \frac{{{\bm {v}_l ^T}{\mathbb E}\left\{ {\bm {y}_{R,B,l}^{}\bm {y}_{R,B,l}^T} \right\}\bm {v}_l }}{{{\bm {v}_l ^T}\bm {v}_l }}{\bm {I}_M}} \right)\bm {v}_l.
\end{align}

Note that in the considered one-ring scatter channel model, the angular spread of each channel is restricted within a narrow region, thus the principal direction of the signal space is oriented towards a specific direction. Let the eigenvalues of $\bm \Xi$ be ${\lambda _1} > {\lambda _2} \ge  \cdots  \ge {\lambda _{2M}} \ge 0$, the principal direction of $\bm \Xi$ is oriented towards a specific direction, which can be characterized by the positive eigenvalue gap ${\lambda _1} - {\lambda _2}$. On each iteration the first coordinate is updated with stepsize $\beta =\kappa/l$, where a $O(1/l)$ rate can be realized by setting $\kappa\ge 1/(2(\lambda _1-\lambda _2))$, then ${\Psi _l}$ can be expected to reach $l^{2\kappa\lambda _1}/(l^{2\kappa\lambda _1}+(2M-1)l^{2\kappa\lambda _2})$. Combining stochastic gradient descent [32] and online learning method [33], the iterative procedure of the principal direction tracking can be described as follow
\begin{align}
\hat{\bm {v} _l} = {\hat {\bm {v}}_{(l-1)}} + {\beta }\left( {\bm {y}_{R,B,l}^{}\bm {y}_{R,B,l}^T - \frac{{{\hat {\bm {v}}_{(l-1)}}\bm {y}_{R,B,l}^{}\bm {y}_{R,B,l}^T{\hat {\bm {v}}_{(l-1)}}}}{{{{\left\| {{\hat {\bm {v}}_{(l-1)}}} \right\|}^2}}}{\bm {I}_M}} \right){\hat {\bm {v}}_{(l-1)}}.
\end{align}

The above iteration can be efficiently carried out by initializing the principal direction $\bm v_l^0$. Firstly, the initialization of can be easily done by setting $\bm v_l^0$ to the first observation that arrives, or to the average of a few received signals. Secondly, the rate of convergence becomes better behaved when the stepsize becomes smaller, which has been widely validated in stochastic gradient descent implementations [32]. Therefore, at time block $l$ upon arrival of a new observation $\bm{y}_{R,B,l}$, we can extract the principal direction $\bm {v}_{l}$ through the principal direction tracking in time without having to re-explore all previously available observations.

\emph{2) Centroid-based Clustering}

The jamming detection is essentially a binary classification problem, and so centroid-based clustering can be used to cluster the received signals into two classes after getting the signal features [34], [35]. Specifically, centroid-based clustering can be efficiently carried out as follows.

We first initialize the centroids and denote the centroids ${\bm \varphi _1}$ and ${\bm \varphi _0}$ as the evens that jamming occurs and does not occur, respectively. We can get ${\bm \varphi _0}$ by using the first observation that arrives, then the current observations will be compared with the prior centroids ${\bm \varphi _0}$ and the following two cases may occur: (i) when the degree of deviation from the current observations to the prior centroids ${\bm \varphi _0}$ is relatively small, it can be judged that there is no jamming and the centroids ${\bm \varphi _0}$ can be updated; (ii) inversely, we can conclude that jamming exists and preserve the prior centroids ${\bm \varphi _0}$, meanwhile, the current observations can be assigned to the centroids ${\bm \varphi _1}$. 

We use $\mathcal C_1$ and $\mathcal C_0$ to denote the existence and absence of a jamming attack, respectively. For the time block $l$, we can obtain the signal features $\hat{\bm v_l}$ by using the principal direction tracking, and we take the ratio of $\hat{\bm t}_{R,l}$ and $\hat{\bm t}_{I,l}$ from $\hat{\bm v_l}$ as a new $M$-dimensional feature $\bm {v}_l^d$. Such density-based feature is very susceptible to interference, so that it can be used for determining whether the jamming exists [35]. By employing the principal direction tracking, the BS can get enough feature samples $\bm {v}_l^d$ from $L$ time blocks, and obtain a feature set ${\mathcal D} \triangleq {\left[ {\bm {v}_1^d,\bm {v}_2^d, \cdots ,\bm {v}_L^d} \right]^T}$. Define the object function of clustering algorithm as ${\mathcal L}\left( {{\bm \varphi _j}} \right)\triangleq {\sum\limits_{{\bm \varphi _j} \in {\mathcal C_j}} {\left\| {{{ {\bm {v}_l^d} }} - {\bm \varphi _j}} \right\|} ^2}$, $j=0,1$, by minimizing the objective function ${\mathcal L}( {\bm \varphi _j} )$, we can obtain robust centroids of two classes. Specifically, these feature samples of the unknown state are put into the classifier as a test set, and the class is judged by $||{\bm {v}_l^d} - {\bm \varphi _0}||/||{\bm {v}_l^d} - {\bm \varphi _1}|| \le \varepsilon$. Note that $\varepsilon$ can be set empirically through simulations. Based on the probability theory and statistics, $\varepsilon$ can be set according to the required false alarm level, for instance, we set $\varepsilon=0.13$ under ${p_{fa}}=5\%$ in this paper. Finally, the jamming detection is completed by obtaining the clustering structure $\bm \varphi _j=\rm {arg}\, min {\mathcal L}( {\bm \varphi _j} )$.

\begin{algorithm}[t]
\caption{OPDAD method for burst jamming detection}
\label{OPDAD}
\begin{algorithmic}[1]
\STATE \textbf{Input:} A sequence of received signal $\bm{y}_{R,B,l}$, $l = {l_0},{l_0} + 1,{l_0} + 2, \cdots ,L$;
\STATE $~~~$\textbf{Set starting time:} Set the clock to time block ${l_0}=1$;
\STATE $~~~$\textbf{Initialization:} Initialize uniformly at random from the unit sphere in ${{\mathbb R}^{2M}}$ and the centroids;
\STATE $~~~$\textbf{Stage 1:} Principal direction tracking.
\STATE $~~~$\textbf{for} $l = {l_0},{l_0} + 1,{l_0} + 2, \cdots ,L$ \textbf{do}
\STATE $~~~~~~$Receive the next observation $\bm{y}_{R,B,l}$;
\STATE $~~~~~~$Update the iterate $\hat{\bm {v}}_{l}$ according to (3);
\STATE $~~~~~~$Update the stepsize;
\STATE $~~~$\textbf{end for}
\STATE $~~~$\textbf{Return:} Density-based feature $\bm {v}_l^d$.

\STATE $~~~$\textbf{Stage 2:} Centroid-based clustering.
\STATE $~~~$Construct the feature set ${\mathcal D} \triangleq {\left[ {\bm {v}_1^d,\bm {v}_2^d, \cdots ,\bm {v}_L^d} \right]^T}$;
\STATE $~~~$\textbf{Repeat:} 
\STATE $~~~~~~$Compute ${\mathcal L}\left( {{\bm \varphi _j}} \right)={\sum\limits_{{\bm \varphi _j} \in {\mathcal C_j}} {\left\| {{{ {\bm {v}_l^d }}} - {\bm \varphi _j}} \right\|} ^2}$, $j=0,1$;
\STATE $~~~~~~$Assign each sample to the nearest cluster centroid according to $\rm min \ {\mathcal L}( {\bm \varphi _j} )$;
\STATE $~~~~~~$Update the centroids ${\bm \varphi _1}$ and ${\bm \varphi _0}$;

\STATE $~~~$\textbf{for} $l = {l_0},{l_0} + 1,{l_0} + 2, \cdots ,L$ \textbf{do}
\STATE $~~~~~~$\textbf{if} $||{\bm {v}_l^d} - {\bm \varphi _0}||/||{\bm {v}_l^d} - {\bm \varphi _1}|| \le \varepsilon$ \textbf{then}
\STATE $~~~~~~$burst jamming does not occur in the $l$th time block;
\STATE $~~~~~~$\textbf{else}
\STATE $~~~~~~$burst jamming occurs in the $l$th time block;
\STATE $~~~$\textbf{end}

\STATE \textbf{Output:} Real time detection result.

\end{algorithmic}
\end{algorithm}

Combining the above two stages, we now summarize the OPDAD method proposed for burst jamming detection in Algorithm 1, we also present the flow chart of the proposed OPDAD method to make it more intuitive. As shown in Fig. 3, the proposed OPDAD method works in an online manner, which can quickly detect the exact attack time block instance by analyzing the newly coming signal. Moreover, unlike the existing methods, our detection method does not rely on the prior knowledge of the attacker, because it only cares the abrupt change in the principal direction of the signal space.

\begin{figure}[t]
 	\centering
 	\includegraphics[width=2.5 in]{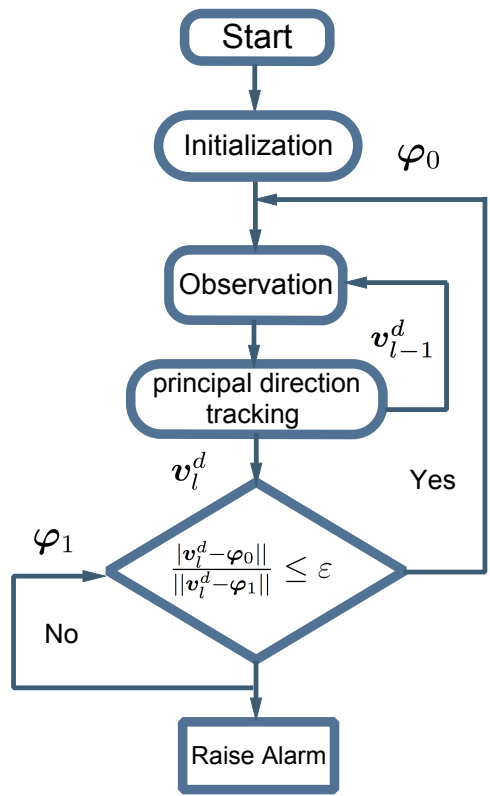}
 	\caption{Flow chart of the proposed OPDAD method.}
 \end{figure}

\section{Performance Analysis}

In this section, the performance of the proposed OPDAD method is analyzed. We provide the convergence rate estimate for the principal direction tracking. Note that the convergence rate is a crucial criterion for such kind of online method to be useful in applications. The convergence rate and the order of convergence of a convergent sequence are quantities that represent how quickly the sequence approaches its limit. We would like to point out that the mathematical expressions of the convergence rate are usually very complicated. After all, the objective of such analysis is usually the expectation of a stochastic nonconvex loss function parameterized by a random variable. We believe that evaluating the performance from the statistical perspective is an interesting issue, which is, however, not suitable for the online method proposed in this paper, and left for future research. Moreover, we derive a nearly optimal finite sample error bound for the proposed OPDAD method.

Before presenting any useful result, we first want to clarify that the received signal $\bm{y}_{R,B,l}$ is a random vector with mean zero and unknow covariance matrix $\bm \Xi$. We assume the eigenvalue gap ${\lambda _1} - {\lambda _2}$ of $\bm \Xi$ is positive. This is a reasonable assumption because the BS equipped with a large number of antennas is elevated at a high altitude, such that there are few surrounding scatters, the angular spread of the legitimate IoT transmitter or the attacker is restricted within a narrow region. Therefore, the principal direction of $\bm \Xi$ is oriented towards a specific direction. Then, we give some basic definitions in preparation for the convergence rate estimate.

\emph{Definition 1 (estimation error): For the $l$th received signal, we define the estimation error of the principal direction as $\tilde {\bm v_l}  \triangleq {\bm {v}_l^\odot} - \hat{\bm v_l}$, where $\bm v_l^\odot$ denotes the true principal direction extracted from the brute force approach, and $\hat{\bm v_l}$ denotes the estimation result of the principal direction through the proposed method.}

Note that we are interested in the angle between $\hat{\bm v_l}$ and $\bm v_l^\odot$, i.e., $\theta ( \hat{\bm v_l},\bm v_l^\odot) = \arccos (\hat{\bm v}_l^T\bm v_l^\odot)$, which plays an important role in the convergence rate result.

\emph{Definition 2 (rescaled iteration index): Define \(L^*_{\beta ,\xi} \triangleq \left\lceil \frac{\xi \log (\lambda _1^{-2}(\lambda _1-\lambda _2)\beta ^{-1})}{-\log \left( 1-\beta (\lambda _1-\lambda _2)\right) }\right\rceil\) as the rescaled iteration index with a tuning parameter \(\xi>0\), where $\beta$ is positive stepsize.}

% where $\beta\triangleq \frac{\log M}{(\lambda _1-\lambda _2)M}$ is an appropriate constant stepsize as defined in [38].

Under the assumption that eigenvalue gap is positive, it was shown that a rescaling of the iteration index \(\left\lceil \frac{\log (\lambda _1^{-2}(\lambda _1-\lambda _2)\beta ^{-1})}{-\log \left( 1-\beta (\lambda _1-\lambda _2)\right) }\right\rceil\) can improve the starting point of stochastic methods for principal direction extraction [36, Lemma 3]. We can extend this definition to introduce a rescaling of the iteration index to a purely streaming setting with a tuning parameter $\xi$, where we only have access to stochastic approximations of principal direction extraction. Note that $L^*_{\beta,\xi}$ increases to infinity as the stepsize $\beta$ decreases to 0. For the presentation of the deterministic initialization, we define the initial iteration index as \(L^0_{\beta,c} \triangleq\left\lceil \frac{\log (cM)}{-\log (1-\beta (\lambda _1-\lambda _2))} \right\rceil\), where $c>0$ is some constant. To obtain the near-optimal convergence rate, we need to choose the stepsize $\beta$ to be inversely proportional to the sample size. Using the same approach and the definition of the initial iteration index $L^0_{\beta,c}$ in [37], the rate of convergence can be obtained under more careful second moment estimates, and convergence results are nearly global in the sense that a randomly selected initial point with the initial iteration index achieves near-optimal convergence rate with high probability.

To prepare for the convergence analysis, we first let the diagonal decomposition of the covariance matrix be $\bm \Xi =\mathbb {E}\left[ {\bm{y}_{R,B,l}}{\bm{y}_{R,B,l}}^T \right] = \bm {V}\bm{\Lambda }\bm {V}^T $, where \(\bm{\Lambda }= \text{ diag }(\lambda _1, \lambda _2, \ldots , \lambda _{2M})\) is a diagonal matrix with diagonal entries \(\lambda _1, \lambda _2, \ldots , \lambda _{2M}\), and \(\bm {V}\) is an orthogonal matrix consisting of column eigenvectors of $\bm \Xi $.

\emph{Definition 3 (rescaled samples, feature space and stepsize): Apply the above orthogonal matrix \(\bm {V}\), we define the rescaled samples as ${\bm z_{R,B,l}}= \bm {V}^T {\bm{y}_{R,B,l}}$, and we have $ \mathbb {E}[{\bm z_{R,B,l}}] = 0$ and $\mathbb {E}\left[ {\bm z_{R,B,l}}{ \bm z_{R,B,l}}^T\right] = \bm{\Lambda }$, then we use $\hat{\bm u_l} = \bm {V}^T \hat{\bm v_l}$ to denote the rescaled feature space. Besides, we define the rescaled stepsize as ${\bar{\beta}} = \lambda _1^2 ( \lambda _1-\lambda _2)^{-1} {\beta }$.}

The principal component of the rescaled random variable \({\bm z_{R,B,l}}\), which we denote by $\bm u_l^\odot$, is equal to \(\bm {e}_1\), where \(\{\bm {e}_1, \ldots , \bm {e}_{2M}\}\) is the canonical basis of \(\mathbb {R}^{2M}\). By applying the linear transformation \(\bm {V}^{T}\) to the stochastic process, we obtain an iterative process $\hat{\bm u_l} = \bm {V}^T \hat{\bm v_l}$ in the rescaled space
\begin{align}
\hat{\bm {u} _l}= {\hat {\bm {u}}_{(l-1)}}+ {\beta }\left( {\bm {z}_{R,B,l}^{}\bm {z}_{R,B,l}^T - \frac{{{\hat {\bm {u}}_{(l-1)}}\bm {z}_{R,B,l}^{}\bm {z}_{R,B,l}^T{\hat {\bm {u}}_{(l-1)}}}}{{{{\left\| {{\hat {\bm {u}}_{(l-1)}}} \right\|}^2}}}{\bm {I}_M}} \right){\hat {\bm {u}}_{(l-1)}}.
\end{align}

Moreover, the angle processes associated with $\hat{\bm {v} _l}$ and $\hat{\bm {u} _l}$ are equivalent, i.e., $\theta ( \hat{\bm v_l},\bm v_l^\odot) =\theta ( \hat{\bm u_l},\bm u_l^\odot) $.

\emph{Definition 4 (ratio of iteration): Define \(r_l^{}\triangleq \hat{\bm v}_l^{}/ \hat{\bm u}_l^{} \) is the ratio of iteration. Geometrically, we observe that the ratio \(r_l^{}\) is the tangent of angle between \(\bm {u}_{l}\) and principal eigenvector \(\bm {u}^* = \bm {e}_1\) after projected onto the two-dimensional subspace spanned by $\hat{\bm u_l}$ and $\bm u_l^\odot$ which are the $1$st and $m$th canonical unit vectors.}

\subsection{Convergence Result and Complexity Analysis}

In this subsection, we present the main convergence results and the complexity analysis of the proposed OPDAD method. To state our convergence results, we present a brief introduction about the notions of convergence and convergence rate. A sequence $\{\bm x_z \}_{z\ge0}\subseteq \mathbb{R}^{Z}$ is said to converge to $\bar {\bm x}\in \mathbb{R}^{Z}$ with rate $\tau_r \in (0,1)$ if $\mathop {\lim }\limits_{z \to \infty } \sup||{\bm {x}_{z+1}} - \bar {\bm x}||/||{\bm {x}_{z}} - \bar {\bm x}|| \le \tau_r$, where $z$ and $\sup$ denote the iteration number and the supremum, respectively. We say $\bm x_z$ converge to $\bar {\bm x}$ R-linearly with rate $\tau^\prime_r \in (0,1)$ if there exists a nonnegative sequence $\{\varepsilon_z \}_{z\ge0}$ such that $||{\bm {x}_{z}} - \bar {\bm x}||\le\varepsilon_z$ for sufficiently large $z$ and $\varepsilon_z \to 0$ linearly with rate $\tau^\prime_r$. Note that we are interested in the angle between $\hat{\bm v_l}$ and $\bm v_l^\odot$, i.e., $\theta ( \hat{\bm v_l},\bm v_l^\odot)$. As with other studies of principal direction extraction [38], [39], we aim to match the convergence rate with the information lower bound. For more details about the minimax information lower bound for estimating the corresponding principal direction, please refer to [38], [39] and references therein. First, for the uniform boundedness and conditions on the stepsize $\beta$ as used in [38], the guarantee of convergence is presented in the following Theorem 1.

\begin{theorem}
Convergence result with deterministic initialization ${\bm v_l^0}$: Define $\mathcal {E}_1$ as an event $\{{\bm v_l^0}: \tan ^2 \theta ( {\bm v_l^0},\bm v_l^\odot)\le c\}$ for some constant $c \in (0,1)$. Suppose the tuning parameter \(\xi\) and the stepsize $\beta$ (defined in Definition 2) satisfy \(M \beta ^{1-2\xi } \rightarrow 0\), $\mathcal {E}_1$ occurs with probability $\mathbb {P}(\mathcal {E}_1) \ge1 - ML^0_{\beta,c}\Gamma - 2ML^*_{\beta ,\xi}\Gamma$ close to 1 even if \(M\rightarrow \infty\), where $L^0_{\beta,c}$ and $L^*_{\beta ,\xi}$ are defined in rescaled iteration index, and $\Gamma \triangleq \exp \left( - c[\lambda _1^2 ( \lambda _1-\lambda _2)^{-1} \beta ]^{-2\xi  } \right)$. Then, the convergence rate of the proposed OPDAD method is satisfied on $\mathcal {E}_1$ with following upper bound
\begin{align} 
\begin{aligned}&\mathbb {E}\left[ \tan ^2 \theta ( \hat{\bm v_l},\bm v_l^\odot)\, |\, \mathcal {E}_1\right] \le \rho^{2(l-L^0_{\beta,c})} + \psi, 
\end{aligned} 
\end{align}
where $\rho \triangleq 1-\beta (\lambda _1 - \lambda _2)<1$ and the positive constant $\psi\triangleq \beta \sum _{m=2}^{2M} \frac{ \lambda _1 \lambda _m + \lambda _1^2 \left[ \lambda _1^2 (\lambda _1-\lambda _2)^{-1} \beta \right] ^{0.5-4\xi } }{\lambda _1-\lambda _m} $ is sufficiently small especially for the high-dimensional data.
\end{theorem}
\emph{Proof:} Please refer to Appendix A. $\hfill\blacksquare$ 

Through matching the convergence rate with the information lower bound [38], [39], Theorem 1 gives a convergence rate estimate for the OPDAD method. Note that the error bound (5) is satisfied on $\mathcal {E}_1$ with probability close to 1 even if \(M\rightarrow \infty\), as long as \(M \beta ^{1-2\xi } \rightarrow 0\) (assuming all other parameters fixed). This means that the convergence rate result is useful in the regime of high-dimensional data analysis. Combining the iterative procedure of the principal direction tracking in (3) with the deterministic initialization in Theorem 1, we can estimate the principal direction by the following steps, same as [38], [39], including repeated filtering, thresholding and orthogonalization. Finally, consistency of an estimator of the whole covariance matrix in spectral norm implies convergence of its principal direction. Next, the convergence result with uniformly randomized initialization is presented in the following Theorem 2.

\begin{theorem}
Convergence result with uniformly randomized initialization ${\bm v_l^0}$: Let ${\bm v_l^0}$ be uniformly sampled from the unit sphere, and define $\mathcal {E}_2$ as an event $\{{\bm v_l^0}: \tan ^2 \theta ( {\bm v_l^0},\bm v_l^\odot)\le \delta\}$ under uniformly randomized initialization with a control factor $\delta \in (0,1/2)$. Suppose the tuning parameter \(\xi\) and the stepsize $\beta$ satisfy $M [\lambda _1^2 ( \lambda _1-\lambda _2)^{-1} \beta ]^{1-2\xi } \le \delta ^2$, there exists a high probability event $\mathcal {E}_2$ with $\mathbb {P}(\mathcal {E}_2) \ge 1 - 2\delta$. Then, with the same high probability, we have 
\begin{align} 
\begin{aligned}&\mathbb {E}\left[ \tan ^2 \theta ( \hat{\bm v_l},\bm v_l^\odot)\, |\, \mathcal {E}_2\right] \le \delta^4 M^2\rho^{2l} + \psi, 
\end{aligned} 
\end{align}
where $\delta^4 M^2 \ll 1$ for high probability event $\mathcal {E}_2$. $\rho<1$ and the small constant $\psi$ are defined in (5).
\end{theorem}

\emph{Proof:} Please refer to Appendix B. $\hfill\blacksquare$ 

Theorem 2 shows that the OPDAD method will converge to a stationary point by using the uniformly random initialization. This is critical because, when $M$ is large, a uniformly distributed initial iterate is nearly perpendicular to the principal component with high probability. Our initial condition allows one to randomly sample ${\bm v_l^0}$ according to a uniform distribution over the sphere, while preserving the near-optimal convergence rate.

\emph{Complexity analysis:} For a $M$-dimensional observation signals, the ED-based method needs to calculate the norm of the received signal, the computing complexity of which is $\mathcal O(M^2)$. As for the SD-based method, it needs to calculate the inverse of a $2M$-dimensional matrix, the computing complexity of which is $\mathcal O(M^3)$. By contrast, the proposed method only requires vector product operations. It has computation complexity $\mathcal O(M)$ per iteration. The proposed method is very easy to implement in practice and can be used as a heuristic method for fast principal component analysis.

\subsection{Finite Sample Analysis}

The batch methods need to store and compute $L$ sample covariance matrix. For comparison, we also give the result of the convergence rate for a fixed sample size. We choose an appropriate stepsize according to the sample size to obtain the explicit estimate of the convergence rate, which refers to the finite sample analysis. Then, when the initial iterate is randomly chosen according to a uniform distribution, the finite sample analysis is presented in the following Theorem 3.

\begin{theorem}
Finite sample error bound: Let ${\bm v_l^0}$ be uniformly sampled from the unit sphere, and define $\mathcal {E}_3$ as an event $\{L: M L^*_{\beta ,\xi} \exp \left( - [\lambda _1^2 ( \lambda _1-\lambda _2)^{-1} \frac{\log L}{(\lambda _1-\lambda _2)L} ]^{-2\xi} \right) \le \delta\}$ with a control factor $\delta \in (0,1/2)$. If the tuning parameter \(\xi\) satisfies $M \left[ \frac{\lambda _1^2}{ ( \lambda _1-\lambda _2)^2} \cdot \frac{ \log L }{ L} \right] ^{1-2\xi} \le \delta ^2$, there exists a high probability event $\mathcal {E}_3$ with $\mathbb {P}(\mathcal {E}_3) \ge 1 - 2\delta$. Then, with the same high probability, the proposed OPDAD method can converge in a finite number of iterations 
\begin{align} \mathbb {E}\left[ \tan ^2 \theta ( \hat{\bm v_l},\bm v_l^\odot)\,|\, \mathcal {E}_3\right] \le \psi',
\end{align}
where $\psi'\triangleq \frac{\lambda _1}{\lambda _1 - \lambda _2} \sum _{m=2}^{2M} \frac{\lambda _m}{\lambda _1 - \lambda _m}\cdot  \frac{\log L}{L}$ becomes sufficiently small as the sample size increases.

\end{theorem}
\emph{Proof:} Please refer to Appendix C. $\hfill\blacksquare$ 

Theorem 3 gives the convergence result of the proposed OPDAD method for a fixed sample size. The finite sample error matches the information lower bound with high probability. Our choice of initial iterate does not require any prior knowledge about the principal component. Therefore, our convergence results are nearly global ($\psi'$ is sufficiently small) in the sense that a randomly selected initial point achieves near-optimal convergence rate with high probability.

Note that we can verify these convergence rate results by the following steps. For the unit sphere ${\mathcal {S}}^{M-1}$, we use the rescaled iteration index as in Definition 2 and the iterative procedure of the principal direction tracking. Consider a partition ${\mathcal {S}}^{M-1} = {\mathcal {S}}_1 \cup {\mathcal {S}}_2$, where ${\mathcal {S}}_1 = \left\{ \bm {v}\in {\mathcal {S}}^{M-1}: |v_1| < 1/\sqrt{2} \right\}$, ${\mathcal {S}}_2 = \left\{ \bm {v}\in {\mathcal {S}}^{M-1}: |v_1| \ge 1/\sqrt{2} \right\}$ and $v_1$ denotes the first coordinate of $\bm {v}$. We refer to ${\mathcal {S}}_1$ and ${\mathcal {S}}_2$ as the cold region and the warm region, respectively. We first focus on the proposed OPDAD method when the initial estimator lies in the warm region ${\mathcal {S}}_2$, which we conveniently call warm start. Such analysis is crucial in obtaining the correct rate of convergence in the proof of Theorems. In terms of the angle $\theta (\bm {v}_l^{0}, \bm {v}^\odot)$ this warm start condition is equivalent to $\theta (\bm {v}_l^{0}, \bm {v}^\odot)\in [0,\pi /4]\cup [3\pi /4,\pi ]$. To avoid uncontrollable variances, we need its first coordinate $v_1$ to be bounded away from 0 throughout the method for $L^*_{\beta ,\xi}$ iterates. We then define an auxiliary region ${\mathcal {S}}_3 = \left\{ \bm {v}\in {\mathcal {S}}^{M-1}: |v_1|\in [1/3, 1]\right\} $ and set the stopping time $\mathcal {N}_w = \inf \left\{ l\ge 0: \bm {v}_l\in {\mathcal {S}}_3^c \right\}$, where $\inf$ denotes the infimum and $\mathcal {A}^c$ for a generic set $\mathcal {A}$ denotes its complement set. Also, for a positive quantity $\Omega$ to be determined later, let $\mathcal {N}_M = \inf \left\{ l\ge 1: \max \left( \max _{1\le l\le 2M} || \bm {y}_{R,B,l}|| ,  |\bm {v}_{(l-1)}\bm {y}_{R,B,l}^T | \right) \ge \Omega^{1/2} \right\}$. In words, $\mathcal {N}_M$ is the first $l$ such that the maximal absolute coordinate of $\bm {y}_{R,B,l}$ exceeds $\Omega^{1/2}$, or the inner product of $\bm {v}_{(l-1)}$ and $\bm {y}_{R,B,l}$, in absolute value, exceeds $\Omega^{1/2}$, whichever occurs earlier. It is convenient to define the rescaled stepsize $ \lambda _1^2(\lambda _1-\lambda _2)^{-1} \beta$. Geometrically, we observe that the ratio iteration is the tangent of angle between $\hat{\bm v_l}$ and $\bm {v}^\odot$. Then combining the initial conditions and tuning parameter (or control factor) in these Theorems, we can verify and match the convergence rate with the information lower bound.

\section{Numerical Results}

In this section, we provide numerical results to illustrate the performance of the proposed detection method. In our simulations, we consider a distance-based path loss model with path loss exponent $\alpha$. The legitimate IoT transmitters and the jammers are uniformly distributed in two circular regions $[30\rm m, 400\rm m]$ and $[100\rm m, 300\rm m]$ with the BS located at the center, respectively. We assume that all legitimate IoT transmitters have the same transmit power, i.e., $P_U=P_{U_k}$ dBm, $k = 1,2, \cdots ,K$, and the jamming power is set to be the same for all attackers, i.e., $P_J=P_{J_n}$, $n = 1,2, \cdots ,N$. We use $P_J^{max}$ to denote the maximum power of the jammers and $N^{max}$ to denote the maximum numbers of the jammers. Besides, the average detection delay as a new performance metric is introduced, it is defined as the extra time blocks needed to discover the occurrence of burst jamming. It is expected that burst jamming can be quickly detected, in other words, the average detection delay is desired to be small. Simulation parameters are listed in Table III.

\begin{table*}[t]\small
\centering
\caption{\small SIMULATION PARAMETERS}
\begin{tabular}{|c|c|} 
\hline
\hline
Parameter & Value \\
\hline
$\alpha$& 3.7  \\
\hline
$L$ & 100  \\
\hline
$P_U$ & 10 dBm  \\
\hline
$t_{g,h}$ & $t_{10,50}$ \\
\hline
$M$ & 64  \\
\hline
$K$ & 10  \\
\hline
$\sigma_B^2$ & -90 dBm \\
\hline
$\varepsilon$ & 0.13 \\
\hline
$P_J^{max}$& 18 dBm \\
\hline
$N^{max}$& 10 \\
\hline
\hline
\end{tabular}

\end{table*}

\subsection{Efficiency of the Proposed OPDAD Method}

To show the efficiency of the proposed OPDAD method, we compare the proposed method with the SD-based method and the SOC-based method in terms of the average time required for obtaining the final solution. The simulation results are provided in Fig. 4. The average simulation times in Fig. 4 are averaged across 1000 random channel realizations using the timing instruction of commercial MATLAB software, i.e., ‘tic’ and ‘toc’. As can be seen, the proposed method is significantly faster than the two competing methods.

\begin{figure}[t]
  \centering
  \includegraphics[width=3.8 in]{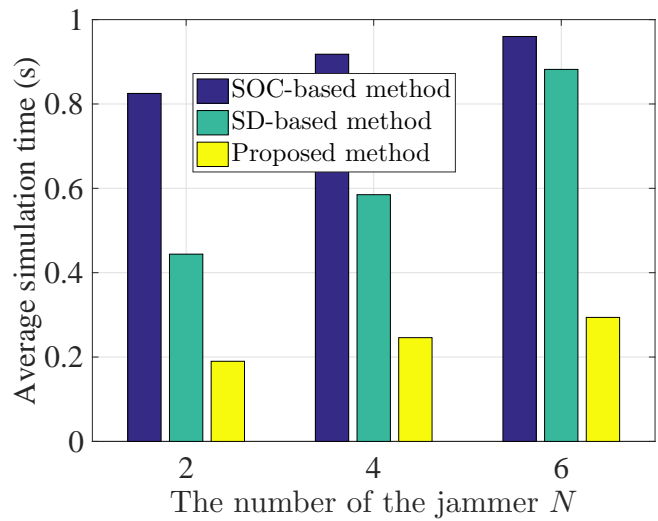}\\
  \caption{The comparison of the average simulation time of different methods, where $P_J=5{\rm{dBm}}, n_r=15$.}\label{f8}
\end{figure}

To verify our convergence rate result is nearly optimal, in Fig. 5, we plot the gaps of the principal direction calculated between the proposed method and the DMF-based method, denoted by $|| \hat{\bm v_l}-\bm v_l^\odot||/||\hat{\bm v_l}||$, for 30 random channel realizations. The DMF-based method transforms a high-dimensional data matrix to a low dimensional projected data to extract the hidden features of the monitoring data matrix. It was shown that the principal space can be found to minimize the information loss between the data points and their projections by such kind of brute force approach. Theoretically, the DMF-based method converges to the globally optimal solution, however, it generally exhibits a high computational complexity. From Fig. 5, we see that the gaps are quite small, i.e., the proposed OPDAD method can get the approximately optimal solution. In fact, despite doing a large number of numerical experiments, we were unable to find a solution with gaps exceeded $10^{-3}$. This indicates that in many cases, the proposed method is expected to find a near-global optimum.

\begin{figure}[t]
  \centering
  \includegraphics[width=3.8 in]{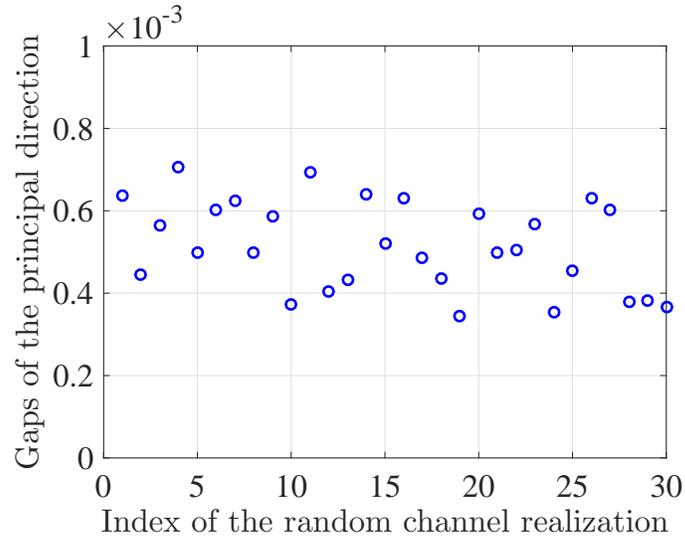}\\
  \caption{The convergence performance of the proposed OPDAD method, where $N=10, P_J=18{\rm{dBm}}, n_r=15$.}\label{f2}
\end{figure}

In Fig. 6, we investigate the quality of the solution achieved by the proposed OPDAD method for small dimensional observations for 3 random channel realizations. To this end, we plot the principal direction angle $\theta ( \hat{\bm v_l},\bm v_l^\odot)$ between the estimation result of the principal direction from the proposed OPDAD method with the DMF-based method. Because the DMF-based method involves the iterative execution of singular value decomposition, it will bring very high computation cost and is not scalable for large-scale problem. So, we mainly focus on the case of small $M$-dimensional observations. Smaller principal direction angle $\theta ( \hat{\bm v_l},\bm v_l^\odot)$ indicates better estimation result of the principal direction from the proposed OPDAD method. As can be observed, the proposed method achieves almost the same estimation results as the brute force approach obtained by the DMF-based method for all considered channel realizations. In conclusion, the proposed OPDAD method is computationally more efficient than the DMF-based method while achieving practically the same performance.

\begin{figure}[t]
  \centering
  \includegraphics[width=3.8 in]{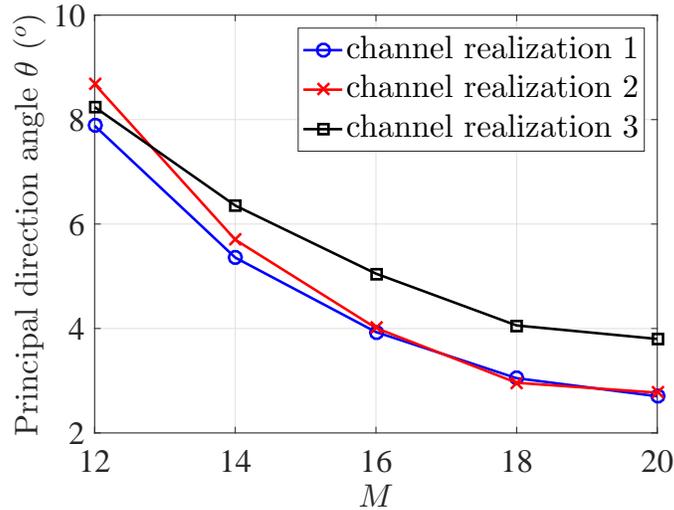}\\
  \caption{The quality of the solution of the proposed OPDAD method, where $K=4, N=2, P_J=5{\rm{dBm}}, n_r=5$.}\label{f3}
\end{figure}

\subsection{Burst Jamming Detectiom}

Fig. 7 plots typical iteration gap curves of the proposed OPDAD method under burst jamming for 3 randomly generated channel realizations. From Fig. 7, we observe that 40 iterations are enough for the OPDAD method to converge for all considered channel realizations. We study the real time performance of the proposed OPDAD detection method under burst jamming. In Fig. 8, we show the average detection delay versus $n_r$ under burst jamming for different $P_J$. It can be seen from Fig. 8 that increasing $n_r$ improves the detection accuracy, i.e., the detection delay becomes smaller with the increase of the number of burst jamming attacks $n_r$. This is because the number of iteration steps required for the OPDAD detection method gets small when $n_r$ increases. Moreover, when the attacking power is high, the detection accuracy increases. This is because higher jamming powers make the attack easier to detect by the BS. It is worth noting that the average detection delay of the proposed OPDAD detection method is much shorter than the ED-based detection method.

\begin{figure}[t]
  \centering
  \includegraphics[width=3.8 in]{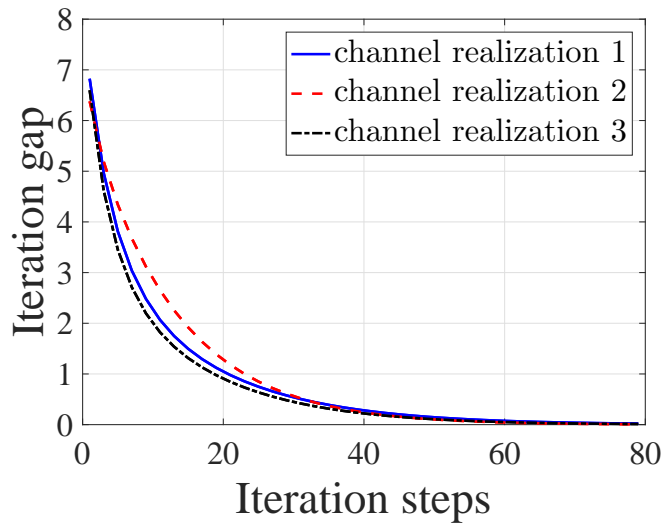}\\
  \caption{The convergence of the OPDAD method under burst jamming, where $N=4, P_J=5{\rm{dBm}}, n_r=15$.}\label{f9}
\end{figure}

\begin{figure}[t]
  \centering
  \includegraphics[width=3.8 in]{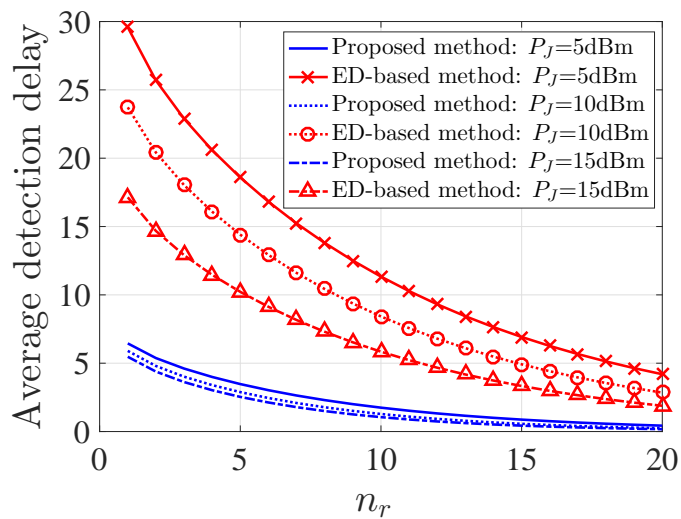}\\
  \caption{The average detection delay versus $n_r$ under burst jamming, where $N=4$.}\label{f5}
\end{figure}

To show the detection performance of the proposed OPDAD detection method, in Fig. 9, we compare the OPDAD detection method under burst jamming with the ED-based detection method in terms of the probability of miss detection. It can be seen from Fig. 9 that the probability of miss detection of the OPDAD method decreases significantly as the transmit power of the attacker increases. From Fig. 9, it is noted that as the number of jammers increases, the attack detection probability also increases. It is worth noting that the probability of miss detection of the proposed OPDAD detection method is much lower than the ED-based detection method. In Fig. 10, we illustrate the probability of miss detection versus the number of jammers under burst jamming for different $n_r$. For all cases, we observe the probability of miss detection decreases with the increasing of $N$. Furthermore, it is observed that the probability of miss detection decreases with increasing of the number of burst jamming attacks $n_r$, which infers that the OPDAD detection method is more effective when the attack frequency is increased.

\begin{figure}[t]
  \centering
  \includegraphics[width=3.8 in]{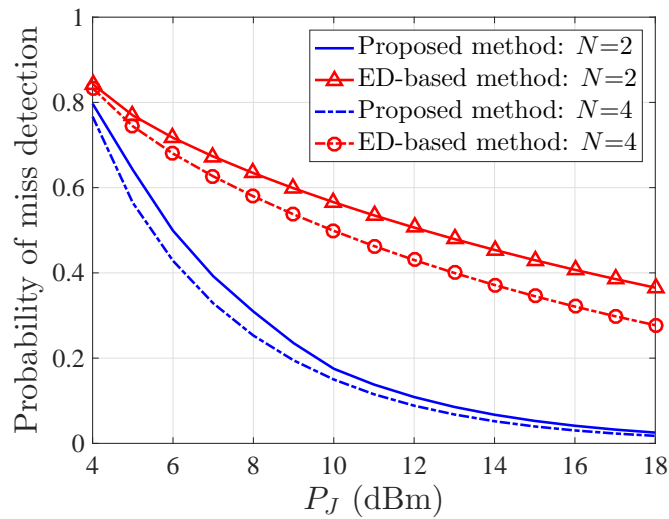}\\
  \caption{The probability of miss detection versus $P_J$ under burst jamming, where $n_r=15$.}\label{f6}
\end{figure}

\begin{figure}[t]
  \centering
  \includegraphics[width=3.8 in]{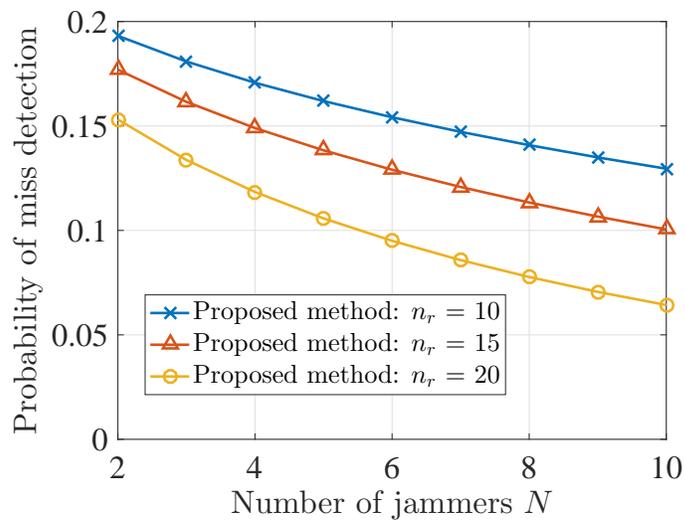}\\
  \caption{The probability of miss detection versus $N$ under burst jamming, where $P_J=10{\rm{dBm}}$.}\label{f7}
\end{figure}

\section{Conclusion}
\label{Sec:Conclusion}

In this paper, we have presented a new kind of online detection method that is possible to detect burst jamming by processing the received signals one by one in a real time manner. The proposed detection method worked in an online manner, at each time a new sample observation was obtained, it decided whether or not burst jamming exists based on all the samples obtained so far. We believe that it is meaningful to investigate how to obtain an approximate algorithm to calculate the principal direction in a close form, which will be one of our future research. Besides, the design of effective defense mechanisms under burst jamming for delay-sensitive IoT applications is also an interesting topic for future research.

\appendix

\subsection{The proof of Theorem 1}
\label{ProofP1}
{
To prove Theorem 1, we first introduce the following proposition.

\emph{Proposition 1 (Proposition 2, [36]): \em Let $T$ be a stopping time with respect to $\{\bm{y}_1,\bm{y}_2,\cdots\}$ such that $\mathbb{P}\{T<\infty\} \leq \frac{1}{\gamma}$ for some  $\gamma\in (1,\infty)$. For $\forall l \geq 1$, let $T^{(l)}$ denote the stopping time obtained by applying $T$ to $\{\bm{y}_l,\bm{y}_{l+1},\cdots\}$, and define $T^* = \min_{l\geq 1} \{ T^{(l)} + l - 1\}$. Then $T^*$ is a stopping time with $\mathbb{E}_{\infty}(T^*)\geq \gamma$ and $\mathcal{D}(T^*)\leq \sup_{l\geq 1}\mathbb{E}_l\left(T^{(l)}\right)$.}

Let $T_c = \inf\{l:l\ge 1,1 - ML^0_{\beta,c} \Gamma - 2ML^*_{\beta ,\xi}\Gamma\geq \eta\}$, where $c>0$, be a stopping time respect to $\{\bm{y}_1,\bm{y}_2,\cdots\}$. Under the condition that $\bm{y}\sim\mathbb{CN}(\bm{0},\bm{I}_M)$, $T_c$ satisfies $\mathbb{P}\{T_c < \infty\} \leq \mathrm{e}^{-\eta}$. Based on the Proposition 4.10 in [37], define $\mathcal{F}_0$ as the hypothesis $\{\bm{y}_l\sim \mathbb{CN}(\bm{0},\bm{I}_M) , \forall l\geq 1\}$ and $\mathcal{F}_1$ as the hypothesis $\{ \bm{y}_l\sim \mathbb{CN}(\bm{0},(1 + s)\bm{I}_M), \forall l\geq 1\}$. $T_c$ is in fact the one-sided sequential probability ratio test that tests $\mathcal{F}_0$ against $\mathcal{F}_1$. We define $\mathcal {E}_{1,1}\triangleq \{ \mathcal {T}_c \le L^0_{\beta,c}\}$ as the event of initial gap and $\mathcal {E}_{1,2}\triangleq \{\sup _{l} \left| r_{(l+ \mathcal {T}_c)}^m \right| \le c\beta \sum _{m=2}^{2M} \frac{ \lambda _1 \lambda _m + \lambda _1^2 \left[ \lambda _1^2 (\lambda _1-\lambda _2)^{-1} \beta \right] ^{0.5-4\xi } }{\lambda _1-\lambda _m}\}$ as the event of iterative increment. By combining these two events together, the event $\mathcal {E}_{1}$ in Theorem 1 can be given by $\mathcal {E}_{1}=  \{ \mathcal {E}_{1,1} \cap \mathcal {E}_{1,2}\}$. In the case of setting the rescaled stepsize, the scaling condition can be derived. By setting the initializing condition, the convergence character of the OPDAD method can be expressed as
\begin{align}
\mathbb {E}\left[ \tan ^2 \theta ( \hat{\bm v_l},\bm v_l^\odot)  \ | \mathcal {E}_1 \right]&\le \left( 1-\beta (\lambda _1 - \lambda _2)\right) ^{l_c}  + \beta\sum _{m=2}^{2M} \frac{  \lambda _{m'}}{\lambda _1 - \lambda _m} , 
\end{align}
where $l_c \triangleq {2(l-L^0_{\beta,c})}$, $\lambda _{m'} \triangleq ( \lambda _1 \lambda _m + \lambda _1^2 \left[ \lambda _1^2 (\lambda _1-\lambda _2)^{-1} \beta \right] ^{0.5-4\xi})$. Then, the high probability event $\mathcal {E}_{1}$ defined in Theorem 1 can be obtained by
\begin{align} 
{\mathbb {P}}(\mathcal {E}_1)\ge 1 - ML^0_{\beta,c} \exp \left( - c \widehat{\beta }^{-2\xi } \right) - 2ML^*_{\beta ,\xi}\exp \left( - c \widehat{\beta }^{-2\xi } \right) , 
\end{align}
% \begin{align} 
% {\mathbb {P}}(\mathcal {E}_1)&\ge \left[ 1 - M L^0_{\beta,c} \exp \left( - c\widehat{\beta }^{-2\xi } \right) \right] \left[ 1- 2ML^*_{\beta ,\xi}\exp \left( - c \widehat{\beta }^{-2\xi } \right) \right] \nonumber \\&\ge 1 - ML^0_{\beta,c} \exp \left( - c \widehat{\beta }^{-2\xi } \right) - 2ML^*_{\beta ,\xi}\exp \left( - c \widehat{\beta }^{-2\xi } \right) , 
% \end{align}
where $\widehat {\beta } \triangleq  \lambda _1^2 ( \lambda _1-\lambda _2)^{-1}\beta$. Combining (8) and (9) completes the proof.

}

\subsection{The proof of Theorem 2}
\label{ProofP2}
{
We now consider the uniform randomized initialization case. First of all, we define $\mathcal {E}_{2,1}$ as the event in which $\bm v_l^0$ is sampled uniformly at random from the unit sphere. Based on Theorem 2 in [38], for a given $\delta >0$, there exists a constant such that $\mathcal {E}_{2,1} \triangleq \{  \tan ^2\theta ( \hat{\bm v_l},\bm v_l^\odot)  \le \delta ^{-2}M\}$. Using the fact about high probability event ${\mathbb {P}}(\mathcal {E}_1) $ in Theorem 1, we have 
\begin{align} 
{\mathbb {P}}(\mathcal {E}_1 \mid \mathcal {E}_{2,1})&\ge 1 - M L^0_{\beta,c_1}  c_1 - 2M L^*_{\beta ,\xi} c_2 \nonumber \\&\ge 1 - 2M L^*_{\beta ,\xi+c_1} c_3 \ge 1-\delta , 
\end{align}
where $c_i \triangleq \exp \left(  - \delta[\lambda _i^2 ( \lambda _1-\lambda _2)^{-1} \beta ]^{-2\xi  }\right)$ for $i=1,2$, \(c_3= {\rm min}(c_1,c_2)\). Define $\mathcal {E}_2\triangleq\mathcal {E}_1 \cap \mathcal {E}_{2,1}$, and we have
\begin{align}
{\mathbb {P}}( \mathcal {E}_1 \cap \mathcal {E}_{2,1}) = {\mathbb {P}}( \mathcal {E}_1 \mid \mathcal {E}_{2,1} ) {\mathbb {P}}( \mathcal {E}_{2,1}) \ge (1 - \delta )^2 \ge 1 - 2\delta.
 \end{align}

Finally, for all \(\xi\) satisfying $M [\lambda _1^2 ( \lambda _1-\lambda _2)^{-1} \beta ]^{1-2\xi } \le \delta ^2$, we have
\begin{align}
\mathbb {E}\left[ \tan ^2 \theta ( \hat{\bm v_l},\bm v_l^\odot) \,|\, \mathcal {E}_2\right] \le c_\delta\left( 1-\beta (\lambda _1 - \lambda _2)\right) ^{2l} +  \beta\sum _{m=2}^{2M} \frac{  \lambda _{m'}}{\lambda _1 - \lambda _m}, 
\end{align}
where $c_\delta(M)\triangleq \delta^4 M^2 \ll 1$ and $\lambda _{m'}$ is defined in (8). Combining the results (11) and (12) together, proof of Theorem 2 is accomplished.

}

\subsection{The proof of Theorem 3}
\label{ProofP3}
{

As the boundary condition for the initial sample interval holds, if, in addition, for the tuning parameter \(\xi\) and control factor \(\delta\) satisfies the condition $M \left[ \frac{\lambda _1^2}{ ( \lambda _1-\lambda _2)^2} \cdot \frac{ \log L }{ L} \right] ^{1-2\xi} \le \delta ^2$, by setting the stepsize in finite sample error analysis $\frac{{\rm{log}}L}{ ( \lambda _1-\lambda _2)L}$ and applying the $L^*_{\beta ,\xi}$ defined in rescaled iteration index, the factor \(\delta\) converges to some absolute constant as long as $M^{-1} \sum _{m=1}^{2M} (\lambda _m / \lambda _1) $ is bounded away from 0 [38]. This is the tightest convergence result known for the principal direction extraction under the near-optimal scaling condition, which yields
\begin{align} 
L^*_{\beta ,\xi} \exp \left( - [\lambda _1^2 ( \lambda _1-\lambda _2)^{-1} \frac{\log L}{(\lambda _1-\lambda _2)L} ]^{-2\xi} \right) \le \delta.
\end{align}

By substituting the uniformly randomized initialization into high probability event ${\mathbb {P}}(\mathcal {E}_2) $ in Theorem 2 and simplifying, there exists a high probability event $\mathcal {E}_3$ with $\mathbb {P}(\mathcal {E}_3) \ge 1 - 2\delta$, where $\mathcal {E}_3$ is defined as $\{L: M L^*_{\beta ,\xi} \exp \left( - [\lambda _1^2 ( \lambda _1-\lambda _2)^{-1} \frac{\log L}{(\lambda _1-\lambda _2)L} ]^{-2\xi} \right) \le \delta\}$ for a fixed sample size. The work by A. Birnbaum et al. [39] studies a different but closely related problem on minimizing the spectral error using a stochastic gradient algorithm. The algorithm’s angular part is equivalent to our online principal direction extraction. Their theoretical guarantees are summarized as: Let ${\bm v_l^0}$ be uniformly sampled from the unit sphere ${\mathcal {S}}^{M-1}$. Given the sample size $L$, by setting $\beta = 16(\lambda _1-\lambda _2) ^{-1} L^{-1}$, the output satisfies with probability at least $3/4$ that
\begin{align} 
\tan ^2 \theta (\hat{\bm v_l},\bm v_l^\odot) \le c_L \cdot \frac{ \lambda _1}{(\lambda _1-\lambda _2)^2 } \cdot \frac{M\log L }{L},
\end{align}
where $c_L$ is an absolute constant for all $L$ sufficiently large. To prove a convergence rate result in terms of the finite sample analysis, we rephrase their main result by reconstruction of $c_L$ via a multiplier $D(2M,L,\delta )$, and establish the minimax information lower bound for estimating the corresponding principal component. The multiplier $D(2M,L,\delta )$ can be computed by
\begin{align} \begin{aligned} D(2M,L,\delta )&= \left( A(\lambda_{1,2})+ \delta \sum _{m=2}^{2M} \frac{\lambda _1\lambda _m + B(\lambda_{1,2}) }{(\lambda _1 - \lambda _m)(\lambda _1 - \lambda _2)} \cdot \frac{\log L}{L} \right) \\&\quad \cdot \left( \frac{\lambda _1}{\lambda _1-\lambda _2}\sum _{m=2}^{2M} \frac{\lambda _m}{\lambda _1-\lambda _m} \cdot \frac{\log L}{L} \right) ^{-1} \\
&= D_1(2M,L,\delta )+ D_2(2M,L,\delta ), \end{aligned} 
\end{align}
where $ A(\lambda_{1,2}) \triangleq \delta ^{2}M \left( 1- {{\beta}} (\lambda _1 - \lambda _2)\right) ^{2l}$, $B(\lambda_{1,2}) \triangleq \lambda _1^2\left[ \lambda _1^2(\lambda _1-\lambda _2)^{-1} {{\beta}}\right] ^{0.5-4\xi }$. The first term on the right hand side of (15) can be expressed by
$$\begin{aligned} D_1(2M,L,\delta )&= A(\lambda_{1,2}) \cdot \left( \frac{\lambda _1}{\lambda _1-\lambda _2}\sum _{m=2}^{2M} \frac{\lambda _m}{\lambda _1-\lambda _m} \cdot \frac{\log L}{L} \right) ^{-1}\\&\le \delta ^{2}M \exp \left( - 2 \cdot \frac{\log L}{L} \right) \cdot \left( \frac{\lambda _1 \lambda _2}{(\lambda _1-\lambda _2)^2} \cdot \frac{\log L}{L} \right) ^{-1}\\&\le  \delta \left[ \frac{\lambda _1^2}{ ( \lambda _1-\lambda _2)^2} \cdot \frac{ \log L }{ L} \right] ^{2\xi -1} L^{-2} \cdot \left( \frac{\log L}{L}\right) ^{-1} \cdot \left( \frac{\lambda _1 \lambda _2}{(\lambda _1-\lambda _2)^2} \right) ^{-1}\\&\le C(\lambda_{1,2}) \frac{ 1 }{ L^{2\xi } \log L ^{2-2\xi } } , \end{aligned}$$
where $ C(\lambda_{1,2}) \triangleq \delta \left( \frac{\lambda _1}{ \lambda _1-\lambda _2} \right) ^{4\xi -2} \left( \frac{\lambda _1 \lambda _2}{(\lambda _1-\lambda _2)^2} \right) ^{-1}$.
The second term on the right hand side of (15) can be expressed by
$$\begin{aligned} D_2 (2M,L,\delta )&= \delta \sum _{m=2}^{2M} \frac{\lambda _1\lambda _m + B(\lambda_{1,2}) }{(\lambda _1 - \lambda _m)(\lambda _1 - \lambda _2)} \cdot \left( \frac{\lambda _1}{\lambda _1-\lambda _2}\sum _{m=2}^{2M} \frac{\lambda _m}{\lambda _1-\lambda _m} \right) ^{-1}\\&= \delta \left( 1 + D(\lambda_{1,2}) E(\lambda_{1,m})\right) . \end{aligned}$$
where $D(\lambda_{1,2}) \triangleq \left[ \lambda _1^2(\lambda _1-\lambda _2)^{-1} {{\beta }}\right] ^{0.5-4\xi }$, $E(\lambda_{1,m}) \triangleq \left[ 2M \left( \sum _{m=2}^{2M} \frac{\lambda _m}{\lambda _1-\lambda _m} \right) ^{-1} + 1 \right]$. Therefore, we have
$$\begin{aligned} D(2M,L,\delta )&= C(\lambda_{1,2}) \frac{ 1 }{ L^{2\xi } \log  L^{2-2\xi } }+ \delta + \delta D(\lambda_{1,2}) E(\lambda_{1,m}), \end{aligned}$$
where $D(2M,L,\delta )$ approaches to some positive constant for high dimensionality samples, in the sense that the finite-sample error matches the minimax information lower bound up to a $\log L$ factor with high probability. Finally, substituing $D(2M,L,\delta )$) into (14) yields the expression (7). This completes the proof.

}

\end{document}